\documentclass[prl,floatfix,twocolumn,showpacs,preprintnumbers,amsmath,amssymb,superscriptaddress,tabularx]{revtex4}
\usepackage{graphicx,float,datetime}

\newcommand{\udt}[3]{#1^{#2}_{\phantom{#2}#3}}
\newcommand{\dut}[3]{#1_{#2}^{\phantom{#2}#3}}

\newcommand{\myf}{\textbf{f}}

\begin{document}

\thispagestyle{empty}

\begin{center}
\title{\large{\bf Generalized Uncertainty Principle in $f\left(R\right)$ Gravity for a Charged Black Hole}}
\date{6$^{\text{th}}$ January 2011}
\author{Jackson Levi Said\footnote{jacksons.levi@gmail.com}}
\affiliation{Physics Department, University of Malta, Msida, Malta}
\author{Kristian Zarb Adami\footnote{kris.za@gmail.com}}
\affiliation{Physics Department, University of Malta, Msida, Malta}
\affiliation{Physics Department, University of Oxford, Oxford, United Kingdom}

\begin{abstract}
{
Using $f\left(R\right)$ gravity in the Palatini formularism, the metric for a charged spherically symmetric black hole is derived, taking the Ricci scalar curvature to be constant. The generalized uncertainty principle is then used to calculate the temperature of the resulting black hole; through this the entropy is found correcting the Bekenstein-Hawking entropy in this case. Using the entropy the tunneling probability and heat capacity are calculated up to the order of the Planck length, which produces an extra factor that becomes important as black holes become small, such as in the case of mini-black holes.
}
\end{abstract}


\maketitle

\end{center}

\section{I. Introduction}
General relativity makes remarkable predictions on the solar scale; however when the galactic and further still the cosmological scale are considered some problems arise. Indeed in 1998 the observation was made that the Universe is in a phase of accelerating expansion \cite{p01} which would require the existence of an exotic form of matter, namely dark energy. It may be that on the cosmological scale new corrections are required. A particular class of alternative theories of gravity that has attracted much attention recently is that of $f\left(R\right)$ gravity, in which the Lagrangian is generalized from $R$ to a function thereof. In particular it has been found that the acceleration may be a result of an additional $R^{-1}$ term in the Lagrangian of the Einstein-Hilbert action \cite{p02}. \newline
As in general relativity the field equations are a result of varying the action for both the gravitational field and the matter field. However in this class of theories there are three ways of proceeding on this front. In general relativity the action is varied with respect to the metric tensor; following the same procedure in the $f\left(R\right)$ model is the first formularism. This paper will focus on the second formularism, called the Palatini formularism, in which the metric tensor and the connection are both considered as independent quantities, and the action is varied with respect to both of them independently. This is done because the field equation would otherwise become unsolvable using traditional methods in the nontrivial cases. It is important to note that the matter fields still remain dependent on the metric tensor in this case and do not have a connection element to them. The last formularism is the metric-affine formularism, where the metric tensor and the affine connection are considered as geometrically separate entities and so the matter Lagrangian will depend on the affine connection in this case. A feature of the resulting field equations in this case is that they reduce to the Einstein field equations of general relativity with a cosmological constant as would be expected for any generalization of a theory.
\newline
It has been shown that $f\left(R\right)$ theories and Einstein's general theory of relativity are equivalent if extra matter fields are incorporated in the Einstein theory \cite{p03}. This is achieved by means of making the so called Einstein conformal transformation $\left(h_{\mu\nu},\phi\right)$ where $\phi$ is an auxiliary scalar field and $h_{\mu\nu}$ is related to the original metric tensor through the conformal transformation $g_{\mu\nu}\rightarrow h_{\mu\nu}=f'\left(R\right)g_{\mu\nu}$, where $h_{\mu\nu}$ will be put in place of the previous metric tensor $g_{\mu\nu}$ in all cases thereafter. However by generalizing the original theory the equations appear in a simpler form than otherwise.
\newline
Since the original derivation by Hawking \cite{p5} of the radiation produced by black holes, there have been numerous extensions and refinements, Refs.\cite{p04,p05,p06}, just to mention a few. However the underlying method is the same; that by starting just behind the horizon, particles may be emitted by means of the fundamental uncertainty in position that results from the quantum mechanical nature of reality. In this way particles may be exited through the vacuum by the horizon, which in turn tunnel through the horizon toward infinity in classically forbidden trajectories. The particles are in effect traveling back in time, since the horizon is locally to the future of the external branch, which is where the particles escape to. The sum of the particles energy must vanish so that energy conservation is preserved in total; thus when a particle escapes to infinity the remaining black hole will lose mass due to the negative energy of the remaining group of particles which were excited out of the vacuum. Now the action for the trajectories of infalling particles will be real since such paths are allowed classically; however the reverse trajectory, that of particles tunneling through the horizon will be complex. In essence the probability of tunneling is based on the imaginary path of this action, or more precisely the probability is an exponential decay of the imaginary part of the trajectory action. There are in general two ways of calculating the imaginary part of the action: the Parikh-Wilczek radial null geodesic method \cite{p07} and the Hamilton-Jacobi method \cite{p08}. These methods are, however, still confined to the semiclassical regime, leaving the quantum gravity problem open.
\newline
It may only be a few years until the production of mini-black holes becomes possible with the LHC at CERN or in other particle accelerators being proposed. Thus it is imperative that the effects of quantum gravity be understood as the black hole mass reduces to Planck dimensions. In particular the significance of gravity in the uncertainty principle is how it will enter into the radiation process. In this way even without a full description of quantum gravity some features may be calculated such as the temperature and eventual entropy of the black hole. It is expected that all the forces including gravity will unify at the Planck scale, which would make this the minimum length scale of the universe; at this scale gravity becomes as important as electroweak and strong interactions.
\newline
In this work we consider the Palatini formularism of gravity and derive the metric for a spherically symmetric charged black hole. In general relativity black holes are classified by a three-parameter family, namely, $\left(M,\,Q,\,a\right)$ or mass, charge and the rotation parameter. In the following analysis a fourth parameter is considered that of a nonvanishing Ricci scalar curvature, which is taken to be constant since otherwise the calculations would be too complex to solve analytically. This is however a cosmological parameter common to the metric describing the whole Universe. We show that the quantum tunneling process is modified when the effects of quantum gravity are taken into account, with the Planck scale as the fundamental scale of nature. We obtain the radiation tunneling probability of this kind of black hole by making a correction to the Hawking-Bekenstein entropy using the generalized uncertainty principle (GUP), in which gravitational effects are included.
\newline
The paper generalizes some of the results offered by Ref.\cite{p4} at a higher order expansion, which does affect the final results, as well as introducing some new quantities that gain importance over the lifetime of the black hole. Furthermore the ultimate fate of the black hole is explored as well as the heat capacity during all the phases of its lifetime.
\newline
The paper uses the signature $\left(-,+,+,+\right)$ and repeated indices are to be summed. Units where $c=1=G=\hbar=k_B$ are used unless explicitly stated otherwise. The paper is organized as follows, in Sec.II the Palatini formularism of $f\left(R\right)$ gravity is introduced. In Sec.III the metric for a spherically symmetric charged black hole is derived using the Palatini formularism and a constant scalar curvature. In Sec.IV the temperature and entropy are calculated followed by the tunneling probability as well as the heat capacity and finally in Sec.V we summarize our results.

\section{II. \protect\myf{}(R) Gravity}
\noindent We start with the action \cite{p10}
\begin{equation}
S=\frac{1}{2\kappa^2}\displaystyle\int\,d^4x\sqrt{-g}\,R+S_{\text{m}}\left(g_{\mu\nu},\psi_{m}\right)
\label{EH_action}
\end{equation}
where $\kappa^2=8\pi$, $g$ is the determinant of the metric $g_{\mu\nu}$ and $S_{\text{m}}$ is a matter action that depends on the curvature of $g_{\mu\nu}$ and the matter fields $\psi_{m}$. The Ricci scalar $R=g^{\mu\nu}R_{\mu\nu}=g^{\mu\nu}\udt{R}{\alpha}{\mu\alpha\nu}$ is all that is needed in the action to produce standard general relativity. Indeed by varying the action Eq.(\ref{EH_action}) with respect to $g_{\mu\nu}$ we find the Einstein field equations
\begin{equation}
R_{\mu\nu}-\frac{1}{2}Rg_{\mu\nu}=\kappa^2 T_{\mu\nu}
\end{equation}
where $T_{\mu\nu}=-\frac{2}{\sqrt{-g}}\frac{\delta S_{\text{m}}}{\delta g^{\mu\nu}}$.
\newline
It was just a few years after the establishment of Einstein's general theory of relativity that modifications were being proposed \cite{p1} and by the people that had previously provided evidence for relativity such as Eddington who had measured the light bending angle during a solar eclipse in 1919. The generalization of Einstein's relativity occurs by making the replacement of $R$ with $f(R)$. It is difficult to find a more general theory, $f\left(R,R^{\mu\nu}R_{\mu\nu},R^{\mu\nu\alpha\beta}R_{\mu\nu\alpha\beta},...\right)$, due to fatal Ostrogradski instabilities that may take hold \cite{p2} in such cases, not that there are not such theories \cite{p9}. The condition for stability is given by $f''>0$, where primes denote differentiation with respect to $R$. Consider the complete action in Palatini gravity \cite{p11}
\begin{equation}
S=\frac{1}{2\kappa^2}\displaystyle\int\,d^4x\sqrt{-g}\,f\left(R\right)+S_{\text{matter}}\left(g_{\mu\nu},\psi_{m}\right)
\end{equation}
where the symbols have the same meaning as in Eq.(\ref{EH_action}). For clarity when dealing with Palatini gravity it is common to replace $R_{\mu\nu}$ with $\mathcal{R}_{\mu\nu}$. Varying the action with respect to $g_{\mu\nu}$ and $\Gamma^{\alpha}_{\mu\nu}$ respectively gives the field equation
\begin{eqnarray}
&&F(\mathcal{R})\mathcal{R}_{\mu\nu}-\frac{f(\mathcal{R})}{2}g_{\mu\nu}=\kappa^2 T_{\mu\nu}\label{metr_var}\\
&&\bar{\nabla}_{\alpha}\left(\sqrt{-g}f'\left(\mathcal{R}\right)g^{\mu\nu}\right)=0
\end{eqnarray}
where $F(\mathcal{R})=f'(\mathcal{R})=\partial f/\partial \mathcal{R}$ and $\bar{\nabla}$ is the covariant derivative defined with the independent connection $\Gamma^{\alpha}_{\mu\nu}$. Taking the trace of Eq.(\ref{metr_var}) yields
\begin{equation}
F\left(\mathcal{R}\right)\mathcal{R}-2f\left(\mathcal{R}\right)=\kappa^2 T
\end{equation}
This equation will be useful in the physical interpretation of the field equations, and, in particular in deriving the metric.

\section{III. Charged Black Holes in Palatini Gravity}
\begin{figure}[H]
\centerline{\includegraphics[width=7cm,height=11cm]{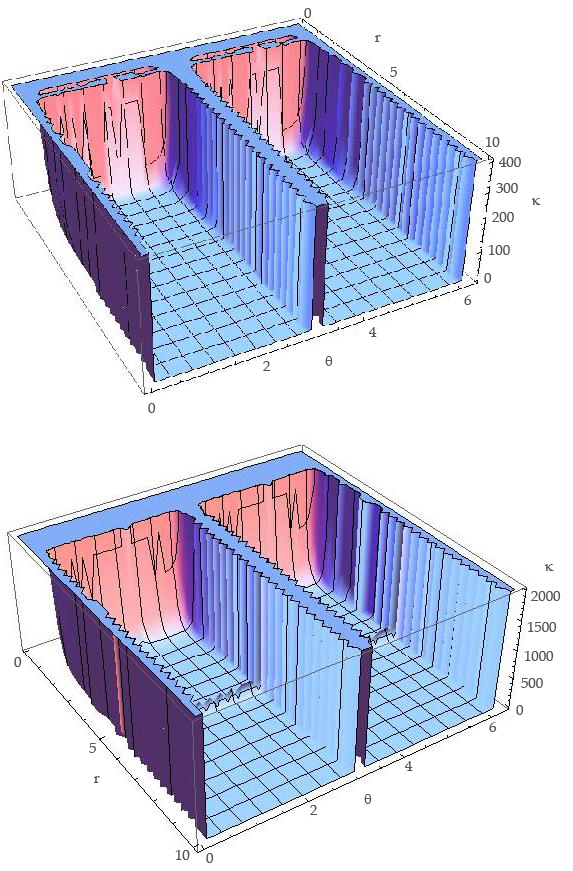}}
\caption{Top: The plot of the Kretschmann scalar for the Reissner-Nordstr\"om black hole with a cosmological constant in terms of the radius and the colatitude angle for unit mass and with $Q=0.8$ and $\Lambda=0.01$. Bottom: The Kretschmann scalar for the Palatini charged black hole with again unit mass such that $Q=0.8$ and with $\mathcal{R}_0=0.01$, $b_1=0.05$, and $R_s=3$ (However, varying this final constant did not make an appreciable difference between the two graphs.)}
\label{kretschmann}
\end{figure}

We consider a spherically symmetric vacuum solution in the coordinate system $\left(t,\,r,\,\theta,\,\phi\right)$ and of the form
\begin{equation}
ds^2=-e^{2\alpha\left(t,r\right)}\,dt^2+e^{2\beta\left(t,r\right)}\,dr^2+r^2\,d\dut{\Omega}{2}{2}
\label{metric_at_infty}
\end{equation}
where
\begin{equation}
d\dut{\Omega}{2}{2}=d\theta^2+\sin^2\theta d\phi^2
\end{equation}
is the line element of the 2-sphere and the usual assumption of vanishing charge is not made. As is well known, the solution remains static \cite{p3} and furthermore by considering two elements of the Ricci tensor it is found that $\alpha$ and $\beta$ are dependent upon each other. In particular it turns out that
\begin{equation}
e^{2\alpha\left(r\right)}R_{rr}+e^{2\beta\left(r\right)}R_{tt}=0
\end{equation}
which in turn implies that
\begin{equation}
\alpha\left(r\right)=-\beta\left(r\right)
\end{equation}
As for the metric tensor, since vanishing charges is not assumed, the electromagnetic stress-energy tensor will have to be included, which is given by
\begin{equation}
T_{\mu\nu}=F_{\mu\rho}\dut{F}{\nu}{\rho}-\frac{1}{4}g_{\mu\nu}F_{\rho\sigma}F^{\rho\sigma}
\end{equation}
where $F_{\mu\nu}$ is the electromagnetic tensor constrained by the Maxwell equations
\begin{align}
\nabla^{\mu}F_{\mu\nu}=0\;\;\;\;\;\;\;\;\;\nabla_{[\mu}F_{\nu\sigma]}=0
\end{align}
Using the metric ansatz and the electromagnetic stress-energy tensor, the only nonvanishing elements of the electromagnetic tensor are found to be
\begin{eqnarray}
&F_{10}=-F_{01}=\left(4\pi\right)^{-1/2}Q_e\,r^{-2}\\
&F_{32}=-F_{23}=\left(4\pi\right)^{-1/2}Q_m\,\sin\theta\,
\end{eqnarray}
where $Q_m$ and $Q_e$ are the total magnetic and electric charge respectively.
\newline
The curvature scalar is restricted to being a constant, i.e. $\mathcal{R}=\mathcal{R}_0$, and
\begin{eqnarray}
\left. f\left(\mathcal{R}\right) \right|_{\mathcal{R}=\mathcal{R}_0}=b_0\\
\left. F\left(\mathcal{R}\right) \right|_{\mathcal{R}=\mathcal{R}_0}=b_1
\end{eqnarray}
This will limit the generality of the obtained solution; however due to the complexity involved an analytic solution will be too difficult to find using traditional techniques. This is still a considerable generalization since a second constant of nature is proposed, namely the definitive of $f\left(\mathcal{R}\right)$; this may also have cosmological consequences that are far different from general relativity \cite{p14,p15}. Considering the (22)-Palatini field equation gives
\begin{eqnarray}
&&b_1\left(e^{-2\left(-\alpha\right)}\left[r\left(-2\alpha_{,r}\right)_1\right]+1\right)-\frac{1}{2}\mathcal{R}_0r^2=\nonumber\\
&&\kappa^2 \left(\frac{Q_m^2}{2r^2\left(4\pi\right)}+\frac{Q_e^2}{2r^2\left(4\pi\right)}\right)
\end{eqnarray}
Letting $Q^2=Q_m^2+Q_e^2$ gives a solution
\begin{equation}
e^{2\alpha}=1-\frac{R_s}{r}+\frac{Q^2}{b_1\,r^2}-r^2\frac{\mathcal{R}_0}{6\,b_1}
\end{equation}
and by considering the limiting case of vanishing charge and Ricci scalar, while letting $F\left(\mathcal{R}\right)\rightarrow1$, gives $R_s\rightarrow2M$, but otherwise this is just a constant.
\newline
The Kretschmann scalar is also given since it reveals the curvature invariant of this metric, and so the inherent distinctions that arise. This scalar is defined by $\mathcal{K}=R_{\mu\nu\sigma\alpha}R^{\mu\nu\sigma\alpha}$, where $R_{\mu\nu\sigma\alpha}$ is the Riemann tensor. Given the immensity of the calculation that is required, this scalar turns out to be remarkably simple, and, in particular, it is given by
\begin{eqnarray}
\mathcal{K}&=&\frac{1}{3 b_1^2 r^8 \left(r \left(6 b_1 (R_s-r)+r^3
   \mathcal{R}_0\right)-6 Q^2\right)^2}\Bigg[2 \nonumber\\
   & &\Big(12 Q^4 r^2 \big(9 b_1^2 \left(28 r^2-102 r R_s+95
   R_s^2\right)+\nonumber\\
   & &3 b_1 r^3 \mathcal{R}_0 (32 R_s-33 r)+26 r^6
   \mathcal{R}_0^2\big)-\nonumber\\
   & &6 Q^2 r^3 \big(-36 b_1^3 \left(3 r^3-16 r^2
   R_s+32 r R_s^2-21 R_s^3\right)\nonumber\\
   & &+6 b_1^2 r^3 \mathcal{R}_0 \left(16 r^2-46 r
   R_s+27 R_s^2\right)+b_1 r^6 \mathcal{R}_0^2 \nonumber\\
   & &(72 R_s-41 r)+2 r^9
   \mathcal{R}_0^3\big)+r^4 \big(54 b_1^4 \big(8 r^4\nonumber\\
   & &-20 r^3 R_s+30 r^2
   R_s^2-32 r R_s^3+15 R_s^4\big)-\nonumber\\
   & &36 b_1^3 r^3 \mathcal{R}_0 \left(11 r^3-24
   r^2 R_s+24 r R_s^2-9 R_s^3\right)+\nonumber\\
   & &9 b_1^2 r^6 \mathcal{R}_0^2 \left(20
   r^2-34 r R_s+21 R_s^2\right)+b_1 r^9 \mathcal{R}_0^3\nonumber\\
   & &(24 R_s-35 r)+3 r^{12}
   \mathcal{R}_0^4\big)-216 Q^6 r\nonumber\\
   & &\left(-27 b_1 r+48 b_1 R_s+2 r^3
   \mathcal{R}_0\right)+\nonumber\\
   & &2 b_1 r^2 \text{cosec} ^2(\theta ) \left(r \left(6
   b_1 (R_s-r)+r^3 \mathcal{R}_0\right)-6 Q^2\right)^2\nonumber\\
   & & \left(3 b_1 r^2 \text{cosec} ^2(\theta )-6 b_1 r R_s+6 Q^2+r^4 (-\mathcal{R}_0)\right)+\nonumber\\
   & &3888 Q^8\Big)\Bigg]
\end{eqnarray}
which is considerably different from the Kretschmann scalar for a charged black hole with a nonzero cosmological constant in standard Einstein-Hilbert gravity. This arises because the above curvature invariant has two cosmological degrees of freedom, while the one considered below has only one. In particular, the metric for the Reissner Nordstr$\ddot{o}$m black hole (Einstein-Hilbert charged black hole) (see Fig.\ref{kretschmann}) is given in Ref.(\cite{p13}) by
\begin{equation}
ds^2=-V\left(r\right)\;dt^2+\frac{dr^2}{V\left(r\right)}+r^2\,d\dut{\Omega}{2}{2}
\end{equation}
where
\begin{equation}
V\left(r\right)=1-\frac{2M}{r}+\frac{Q^2}{r^2}-\frac{1}{3}\Lambda r^2
\end{equation}
which results in a scalar invariant given by
\begin{eqnarray}
\mathcal{K}_{GR}&=&\frac{1}{3 r^6 \left(6 M-3 \left(Q^2+1\right) r+r^3 \Lambda
   \right)^2}\Bigg[4\nonumber\\
   & &\Bigg(3 r^6 \Lambda ^2 \bigg(126 M^2-6 M \left(20 Q^2+17\right)\nonumber\\
   & &r+\left(37 Q^4+65 Q^2+30\right) r^2\bigg)-\nonumber\\
   & &9 r^3 \Lambda  \big(-72 M^3+12
   M^2 \left(11 Q^2+8\right) r\nonumber\\
   & &-4 M \left(19 Q^4+29 Q^2+12\right)
   r^2+\nonumber\\
   & &\left(Q^2+1\right) \left(16 Q^4+23 Q^2+11\right) r^3\big)+\nonumber\\
   & &27\Bigg(60 M^4-8 M^3 \left(11 Q^2+8\right) r+\nonumber\\
   & &2 M^2 \left(29 Q^4+40
   Q^2+15\right) r^2-2 M\nonumber\\
   & &\left(Q^2+1\right) \left(10 Q^4+11 Q^2+5\right)
   r^3+\nonumber\\
   & &\left(Q^2+1\right)^2 \left(3 \left(Q^4+Q^2\right)+2\right)
   r^4\Bigg)+\nonumber\\
   & &r^9 \Lambda ^3 \left(48 M-\left(38 Q^2+35\right) r\right)+\nonumber\\
   & &r\text{cosec}^2(\theta ) \left(6 M-3 \left(Q^2+1\right) r+r^3 \Lambda \right)^2\nonumber\\
   & &\bigg(-12 M+6 Q^2 r-2 r^3 \Lambda +\nonumber\\
   & &3 r \text{cosec}^2(\theta )\bigg)+6 r^{12}
   \Lambda ^4\Bigg)\Bigg]
\end{eqnarray}

\begin{figure}[H]
\centerline{\includegraphics[width=7.5cm,height=6.5cm]{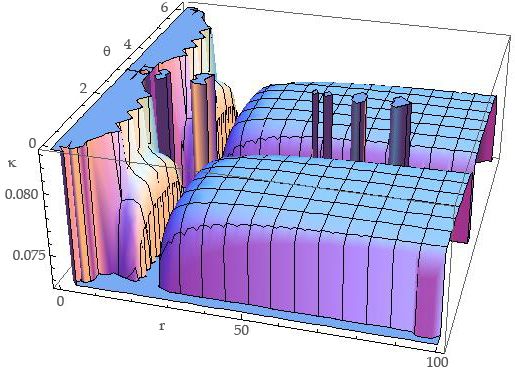}}
\caption{A qualitative comparison between a charged black hole in general relativity with a nonzero cosmological constant and a Palatini charged black hole with a nonzero cosmological constant. The figure shows that on the small scale there is a fine difference between the two metrics, and furthermore the flat planes show where significant differences arise.}
\label{compar}
\end{figure}

Finally we give a graphical comparison which shows that this degree of freedom actually has a significant effect on the background spacetime as shown in Fig.(\ref{compar}) even for small sample values
A look at Fig.(\ref{compar}) shows that quantitative differences arise where there are flat planes, signifying large differences between the two metrics, and the fine structural differences, shown by the curved parts of the graph.
\newline
In the next section, for completeness, the case of a falling observer will be considered for the metric in Eq.(\ref{metric_at_infty}).

\subsection{The Infalling Observer}
The metric in Eq.(\ref{metric_at_infty}) contains two singularities which implies that measurements cannot be made across such bounds using classical observers since measurements in this metric are made by an observer at infinity. However, when a falling observer is considered, such branches of spacetime become attainable. To derive such a metric, tortoise coordinates are considered, first by defining a new radial variable, $r^\ast$, by
\begin{equation}
\frac{dr^{\ast}}{dr}=e^{-2\alpha}
\label{tortoise}
\end{equation}
and then by considering a new time coordinate, which can be interpreted as the new falling time coordinate
\begin{equation}
v=t\pm\left(r-r^{\ast}\right)
\end{equation}
where the positive (upper) transformation refers to an ingoing observer and the negative (lower) transformation refers to the outgoing observer. Combining this transformation with Eq.(\ref{tortoise}) gives the final metric
\begin{eqnarray}
ds^2&=&-\left(1-\frac{R_s}{r}+\frac{Q^2}{b_1\,r^2}-r^2\frac{\mathcal{R}_0}{6\,b_1}\right)dv^2-\nonumber\\
& &\left(-1-\frac{R_s}{r}+\frac{Q^2}{b_1\,r^2}-r^2\frac{\mathcal{R}_0}{6\,b_1}\right)dr^2\pm\nonumber\\
& &2\left(\frac{R_s}{r}-\frac{Q^2}{b_1\,r^2}+r^2\frac{\mathcal{R}_0}{6\,b_1}\right)\,dv\,dr+r^2\,d\dut{\Omega}{2}{2}\nonumber\\
& &
\end{eqnarray}
This metric is significant because it does not contain the singularities that the metric in Eq.(\ref{metric_at_infty}) has; however, it contains the cross term between the radial and the timelike coordinate, meaning that the observer must be either falling in or out over time.

\section{IV. Black Hole Lifetimes}
\subsection{Entropy}
Now that a charged solution to the Palatini formularism has been found, we can move on to calculating the entropy of this black hole. In particular, we will employ some of the calculation techniques used in Ref.\cite{p4}, that is, we will use the generalized uncertainty principle in lieu of the uncertainty principle when deriving the Hawking temperature, which will turn out to predict a different fate for black holes. The radii of the horizons will first be derived. The roots of
\begin{equation}
e^{2\alpha}=0
\label{hor_eqn}
\end{equation}
will give the radii of the inner and outer horizons. However, there will also be two others roots from Eq.(\ref{hor_eqn}), i.e. there will be four roots, which can be classified by
\begin{equation}
r_1>r_2>r_3>r_4
\label{hor_condi}
\end{equation}
where $r_1$ will correspond to the cosmological horizon and $r_4$ will turn out to be negative and so not a physical manifestation of the black hole and so not measurable by observers at infinity. The negativity of $r_4$ emerges from the lack of cubic term in Eq.(\ref{hor_eqn}), and the cosmological horizon emerges out of the fact that the Universe had a beginning. Now the inner (or {\it Cauchy}) and outer horizons will be represented by $r_-$ and $r_+$, respectively. The radii that will represent the horizons given by Eq.(\ref{hor_eqn}) and satisfying the condition given by Eq.(\ref{hor_condi}) are given by
\begin{eqnarray}
&&r_+=-\frac{\sqrt{Y}}{2}+\frac{1}{2} \sqrt{\frac{12b_1}{\mathcal{R}_0}-Y+\frac{24Mb_1}{\mathcal{R}_0\sqrt{Y}}}\\
&&r_-=\frac{\sqrt{Y}}{2}-\frac{1}{2} \sqrt{\frac{12b_1}{\mathcal{R}_0}-Y-\frac{24Mb_1}{\mathcal{R}_0\sqrt{Y}}}
\end{eqnarray}
where
\begin{eqnarray}
X&&=\Bigg(-\frac{432 b_1^3}{\mathcal{R}_0^3}\nonumber\\
&&+\sqrt{\left(-\frac{432 b_1^3}{\mathcal{R}_0^3}-\frac{2592b_1Q^2}{\mathcal{R}_0}+\frac{3888M^2b_1^2}{\mathcal{R}_0}\right)^2}\nonumber\\
&&\overline{-4 \left(\frac{36b_1^2}{\mathcal{R}_0^2}-\frac{72Q^2}{\mathcal{R}_0}\right)^3}\nonumber\\
&&-\frac{2592b_1Q^2}{\mathcal{R}_0^2}+\frac{3888M^2b_1^2}{\mathcal{R}_0^2}\Bigg)^{1/3}\\
Y&&=\frac{\sqrt[3]{2} \left(\frac{36b_1^2}{\mathcal{R}_0^2}-\frac{72Q^2}{\mathcal{R}_0}\right)}{3 X}+\frac{4b_1}{\mathcal{R}_0}+\frac{X}{3 \sqrt[3]{2}}
\end{eqnarray}
As in Ref.\cite{p4} we consider the black hole as a cube of side length twice the outer horizon radius; the uncertainty in position of a Hawking particle is then given by
\begin{eqnarray}
&&\Delta x=2r_+\nonumber\\
&&=2\left(-\frac{\sqrt{Y}}{2}+\frac{1}{2} \sqrt{\frac{12b_1}{\mathcal{R}_0}-Y+\frac{24Mb_1}{\mathcal{R}_0\sqrt{Y}}}\right)\nonumber\\
&&=-\sqrt{Y}+\sqrt{\frac{12b_1}{\mathcal{R}_0}-Y+\frac{24Mb_1}{\mathcal{R}_0\sqrt{Y}}}
\label{pos_uncer}
\end{eqnarray}
Applying the uncertainty principle, in its usual form, to the energy of the Hawking particles being emitted
\begin{eqnarray}
&&\Delta E\approx c\Delta p\approx\frac{\hbar c}{\Delta x}\nonumber\\
&&=\hbar c\left[-\sqrt{Y}+\sqrt{\frac{12b_1}{\mathcal{R}_0}-Y+\frac{24Mb_1}{\mathcal{R}_0\sqrt{Y}}}\right]^{-1}
\end{eqnarray}
It is then straightforward to calculate the temperature \cite{p5}. Indeed it is related to the outer horizon radius by
\begin{equation}
T=\frac{1}{4\pi r_+}=\frac{1}{2\pi\Delta x}
\label{hawk_temp}
\end{equation}
Now the Bekenstein-Hawking entropy, $S$, will be related to the black hole mass by
\begin{equation}
T=\frac{dE}{dS}=\frac{dM}{dS}
\label{hawk_beken_temp}
\end{equation}
since geometric units are being used. Thus using Eq.(\ref{pos_uncer}), Eq.(\ref{hawk_temp}) and Eq.(\ref{hawk_beken_temp})
\begin{eqnarray}
&&S\left(M\right)=A_2\left(-\sqrt{Y}+\sqrt{\frac{12b_1}{\mathcal{R}_0}-Y+\frac{24Mb_1}{\mathcal{R}_0\sqrt{Y}}}\right)^{-2}\nonumber\\
&&\times\displaystyle\int dM\left[-\sqrt{Y}+\sqrt{\frac{12b_1}{\mathcal{R}_0}-Y+\frac{24Mb_1}{\mathcal{R}_0\sqrt{Y}}}\right]\nonumber\\
& &
\label{can_entropy}
\end{eqnarray}
where $A_2=4\pi r_+^2$ is the surface area of the black hole outer horizon.
\newline
Because of the extreme nature of the event horizon of such black holes, terms that are negligible in other physical processes can take hold and make a significant contribution. In this case the interaction of gravity is considered in the emission process, in particular the GUP is considered \cite{p6,p7} where
\begin{equation}
\Delta x\geq\frac{\hbar}{\Delta p}+\alpha^2L_p^2\frac{\Delta p}{\hbar}
\label{gen_uncer}
\end{equation}
where $L_p=\sqrt{\frac{G\hbar}{c^3}}$ is the Planck length and $\alpha$ is a constant, normally set to the order of unity but which in string theory is found to correspond to the string tension. The second term in Eq.(\ref{gen_uncer}) relates to the uncertainties due to the gravitational effects, and so they will only become significant when $\Delta x\approx L_p$. Considering again $\Delta x$ as $2r_+$ will result in a range of possible values for $\Delta p$ by Eq.(\ref{gen_uncer}), such that
\begin{eqnarray}
\frac{r_+}{\alpha^2\dut{L}{p}{2}}& &\left(1-\sqrt{1-\frac{\alpha^2 \dut{L}{p}{2}}{\dut{r}{+}{2}}}\right)\leq\frac{\Delta p}{\hbar}\nonumber\\
& &\leq\frac{r_+}{\alpha^2\dut{L}{p}{2}}\left(1+\sqrt{1-\frac{\alpha^2\dut{L}{p}{2}}{\dut{r}{+}{2}}}\right)
\label{momentum_res}
\end{eqnarray}
Taking a series expansion of the first inequality in Eq.(\ref{momentum_res}) and taking the lower limit gives
\begin{eqnarray}
\frac{\Delta p}{\hbar}& &=\frac{r_+}{\alpha^2\dut{L}{p}{2}}\left(1-\sqrt{1-\frac{\alpha^2\dut{L}{p}{2}}{\dut{r}{+}{2}}}\right)\nonumber\\
& &=\frac{1}{2r_+}\left(1+\frac{\alpha^2\dut{L}{p}{2}}{4\dut{r}{+}{2}}+\frac{\alpha^4\dut{L}{p}{4}}{8\dut{r}{+}{4}}\right)\nonumber\\
& &=\frac{1}{\Delta x}\left(1+\frac{\alpha^2\dut{L}{p}{2}}{\left(\Delta x\right)^2}+2\frac{\alpha^4\dut{L}{p}{4}}{\left(\Delta x\right)^4}\right)+\mathcal{O}\left(\dut{L}{p}{5}\right)\nonumber\\
& &
\end{eqnarray}
Substituting into Eq.(\ref{gen_uncer})
\begin{eqnarray}
\Delta x'& &=\Delta x\Bigg[\left(1+\frac{\alpha^2\dut{L}{p}{2}}{\left(\Delta x\right)^2}+2\frac{\alpha^4\dut{L}{p}{4}}{\left(\Delta x\right)^4}\right)^{-1}+\nonumber\\
& &\frac{\alpha^2\dut{L}{p}{2}}{\left(\Delta x\right)^2}\left(1+\frac{\alpha^2\dut{L}{p}{2}}{\left(\Delta x\right)^2}\right)\Bigg]
\end{eqnarray}
Finally the resulting corrected Hawking temperature becomes
\begin{eqnarray}
&T'&=\frac{1}{2\pi\Delta x'}\nonumber\\
& &=T\Bigg[\left(1+\frac{\alpha^2\dut{L}{p}{2}}{\left(\Delta x\right)^2}+2\frac{\alpha^4\dut{L}{p}{4}}{\left(\Delta x\right)^4}\right)^{-1}+\nonumber\\
& &\frac{\alpha^2\dut{L}{p}{2}}{\left(\Delta x\right)^2}\left(1+\frac{\alpha^2\dut{L}{p}{2}}{\left(\Delta x\right)^2}\right)\Bigg]^{-1}\nonumber\\
& &
\label{gen_temp}
\end{eqnarray}
Similar to Eq.(\ref{can_entropy}) the corrected entropy can be calculated analogously, leading to
\begin{eqnarray}
S'\left(M\right)&=&S\left(M\right)-2\alpha^4\dut{L}{p}{4}B\left(M\right)+\mathcal{O}\left(\dut{L}{p}{5}\right)\nonumber\\
\label{mod_ent}
\end{eqnarray}
where $S\left(M\right)$ is given by Eq.(\ref{can_entropy}) and
\begin{eqnarray}
& &B\left(M\right)=\displaystyle\int dM\nonumber\\
&&\left[A_2\left(-\sqrt{Y}+\sqrt{\frac{12b_1}{\mathcal{R}_0}-Y+\frac{24Mb_1}{\mathcal{R}_0\sqrt{Y}}}\right)^{-5}\right]\nonumber\\
&&
\end{eqnarray}
The Planck correction term becomes fourth order from second order as in Ref.\cite{p4} due to the expansion in temperature in Eq.(\ref{gen_temp}) being taken to fourth order in Eq.(\ref{mod_ent}). That is, the second order term cancels in this case. Thus the entropy is modified by a corrected temperature given by the GUP. This still returns only values that have statistical significance when Planck dimensions are reached by the horizon, and indeed it does affect the final state, as will be shown.

\subsection{Tunneling Probability}
In classical physics any units of mass that enter a black hole cannot escape, that is, black holes are perfect absorbers. However, the horizon is not a classical surface, in that since it is a one-way membrane, it may excite particles from the vacuum such that they may tunnel through, and for those with enough energy, they may escape to infinity. Thus, despite there being no such classical trajectory, there still is a way for black holes to emit through classical forbidden regions, by considering the fundamental quantum nature of reality. The process of emission is hence a semiclassical one, since the event horizon is a result of classical physics, and emission through it is a quantum process. Using the WKB approximation \cite{p12}, the tunneling probability is a function of only the imaginary part of the classical action of the trajectory $I$, namely,
\begin{equation}
\Gamma\sim e^{-2I_m\left(I\right)}
\end{equation}
where $I_m\left(I\right)=\text{Im}\left(I\right)$. This can be represented, as in Ref.\cite{p8}, as
\begin{equation}
\Gamma\sim\frac{e^{S_f}}{e^{S_i}}=e^{\Delta S}
\label{can_prob}
\end{equation}
where $\Delta S$ is the difference between the final and initial entropies of the black hole.
\newline
Because of the correction in the Bekenstein-Hawking entropy in Eq.(\ref{mod_ent}), the change in entropy will turn out to be
\begin{equation}
\Delta S'=\Delta S-2\alpha^4\dut{L}{p}{4}\Delta B
\label{corr_entro}
\end{equation}
where
\begin{eqnarray}
& &\Delta S=S\left(M-E\right)-S\left(M\right)\\
& &\Delta B=B\left(M-E\right)-B\left(M\right)
\end{eqnarray}
and $E$ is the energy of a particle being emitted. Now substituting Eq.(\ref{corr_entro}) into Eq.(\ref{can_prob}) gives
\begin{equation}
\Gamma'\sim\Gamma\;Exp\left[\alpha^4\dut{L}{p}{4}\Delta B\right]
\end{equation}
which corrects the tunneling probability up to the Planck length. This indicates that as the black hole reduces in size to Planck dimensions quantum gravity takes hold, and an exponential factor drastically modifies the emission rate of particles through the horizon membrane. The effect of the extra factor will be minimal for very large black holes; however, the generalized tunneling probability will take effect when considering black holes at the end of their evaporation process.

\subsection{Heat Capacity}
Using the semiclassical approach, the heat capacity of a black hole may be calculated by first giving the inverse temperature, which is given by
\begin{equation}
\beta=T^{-1}=\frac{dS}{dM}
\end{equation}
The heat capacity is then given by $C=\frac{dM}{dT}$, which when using Eq.(\ref{hawk_temp}) results in
\begin{eqnarray}
& &C=\frac{1}{\pi T^2}\Bigg[\frac{1}{\sqrt{Y}}+\nonumber\\
& &\left(\frac{12b_1}{\mathcal{R}_0}-Y+\frac{24Mb_1}{\mathcal{R}_0\sqrt{Y}}\right)^{-1/2}\left(1+\frac{12Mb_1}{\mathcal{R}_0Y^{3/2}}\right)\nonumber\\
& &\Bigg]^{-1}\frac{1}{W}\nonumber\\
& &
\label{uncer_heat}
\end{eqnarray}
where
\begin{equation}
W=\frac{dY}{dM}=\left[\frac{\sqrt[3]{2}\left(\frac{72Q^2}{\mathcal{R}_0}-\frac{36\dut{b}{1}{2}}{\dut{\mathcal{R}}{0}{2}}\right)}{3X^2}+\frac{1}{3\sqrt[3]{2}}\right]Z
\end{equation}
and
\begin{eqnarray}
Z& &=\frac{dX}{dM}\nonumber\\
& &=\frac{1}{3}X^{-2}\Bigg[\frac{1}{2}\left(X^3+\frac{432\dut{b}{1}{3}}{\dut{\mathcal{R}}{0}{3}}+\frac{2592b_1Q^2}
{\dut{\mathcal{R}}{0}{2}}\right)^{-1}\nonumber\\
& &\Big(2\left(-\frac{432\dut{b}{1}{3}}{\dut{\mathcal{R}}{0}{3}}-\frac{2592b_1Q^2}{\mathcal{R}_0}
+\frac{3888M^2\dut{b}{1}{2}}{\mathcal{R}_0}\right)\nonumber\\
& &\frac{7776M^2\dut{b}{1}{2}}{\mathcal{R}_0}\Big)
+\frac{7776M\dut{b}{1}{2}}{\dut{\mathcal{R}}{0}{2}}\Bigg]
\end{eqnarray}
However when the GUP is incorporated the heat capacity becomes
\begin{eqnarray}
& &C'=\nonumber\\
& &C\left[1-\frac{\alpha^4\dut{L}{p}{4}}{\left(-\sqrt{Y}+\sqrt{\frac{12b_1}{\mathcal{R}_0}-Y+\frac{24Mb_1}{\mathcal{R}_0\sqrt{Y}}}\right)^4}\right]^2\nonumber\\
& &\Bigg[1+
\frac{\alpha^4\dut{L}{p}{4}}{\left(-\sqrt{Y}+\sqrt{\frac{12b_1}{\mathcal{R}_0}-Y+\frac{24Mb_1}{\mathcal{R}_0\sqrt{Y}}}\right)^4}\nonumber\\
& &\left(\frac{1}{8\pi^5}-1\right)\Bigg]^{-1}\nonumber\\
& &
\label{gup_heat}
\end{eqnarray}
in which Eq.(\ref{gen_temp}) was used.
\newline
When $L_p$ vanishes Eq.(\ref{gup_heat}) reduces to Eq.(\ref{uncer_heat}) as would be expected. The effects of the generalization will be given in Sec. V along with the consequences of the other generalizations.

\subsection{The Black Hole Remnant}
As the mass of a black hole with a background metric given by Eq.(\ref{metric_at_infty}) vanishes, the heat capacity also vanishes when the standard uncertainty principle is employed; however, when one incorporates the GUP into the theory, the heat capacity suddenly vanishes outside of the singularity at a critical mass, say, $M_c$. This mass depends heavily on the cosmological parameters $\mathcal{R}_0$ and $b_1$; however it is quite tediously long. The point to take from this is that the size of the remnant black hole depends very strongly on the large scale parameters of the Universe in this case. This indicates as in Refs.\cite{p16,p17} that the black hole enters a new phase that does not include evaporation, and thus a black hole remnant would emerge which would truly be a remnant since most thermodynamical change would have ceased, leaving just the classical singularity and the horizons behind.
\begin{figure}[H]
\centerline{\includegraphics[width=6.0cm,height=5.5cm]{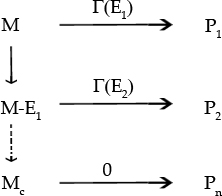}}
\caption{The process by which the black hole emits a series of particles $P_i$ with energy $E_i$ with a probability of emission $\Gamma\left(E_i\right)$. However, after a certain number of particles, say, $n$ particles, has been emitted, the black hole stops emitting particles, and so the probability of emission vanishes, which in turn leaves a black hole remnant behind with a critical mass $M_c$.}
\label{rem}
\end{figure}
The process will follow as shown in Fig.(\ref{rem}) in which particles will be emitted with energies $E_1,\;E_2,\;...$, up to the point where emission ceases.
\newline
In this way as the black attains Planck dimensions, its entropy tends to infinity. Given that the entropy increases in this way, the mass cannot decrease to zero unless a new quantum gravity process allows for the dissipation of the new degrees of freedom produced by the previous process of evaporation.

\section{V. Conclusion}
Through the Palatini formularism of $f\left(\mathcal{R}\right)$ gravity, the metric for a spherically symmetric charged mass was derived for a constant Ricci scalar. This turned out to be the regular de Sitter metric for the Reissner-Nordstr\"om metric with some extra and significant factors. These factors were found to be strongly dependent on the type of $f\left(\mathcal{R}\right)$ gravity being employed. The metric introduced in this work thus generalizes the general relativistic one; however, a further degree of freedom is allowed, namely, the derivative of the $f\left(\mathcal{R}\right)$ function with respect to the Ricci scalar with a constant curvature scalar, which was shown to give order or magnitude differences in the background curvature, which can be seen by comparing Fig.(\ref{kretschmann}) with Fig.(\ref{compar}).
\newline
For the second part of this paper, the tunneling probability for the derived black hole was calculated. The uncertainty principle with relation to the position of the particle was generalized to encompass gravitational effects that take effect on Planck scales, but still have some minor effect on large black holes. In using the GUP the temperature and thus the entropy was calculated correcting the Bekenstein-Hawking result. Using this entropy result, the WKB semiclassical approximation was applied, and through it the tunneling probability was calculated for the generalized metric, thus showing a difference between solutions in $f\left(\mathcal{R}\right)$ theories and those with this function being taken as unity. This could potentially prove vital in differentiating between competing theories of gravity if mini-black holes are produced in particle accelerators. It is in these types of black holes that this tunneling probability takes on significant practical importance. It may thus be possible in just a few years to determine the $f\left(\mathcal{R}\right)$ function and whether it is actually not unity in reality.
\newline
Finally it was shown how a black hole of this nature would proceed to attain Planck scale dimensions and it turned out that the evaporation process described by Hawking \cite{p5} cannot make the mass vanish in this phase of the black hole lifetime due to the gravitational effects of the GUP. Lastly the heat capacity was calculated for both the standard uncertainty principle and the generalized uncertainty principle, giving different masses where this parameter vanishes. This quantity is an important factor in black hole thermodynamics because it is inextricably related to the entropy and so to the information content of a black hole, since $\Delta S=-\Delta I$, where $I$ is the information content of the black hole. At the time of publication further work was being conducted by Moon in Ref.\cite{p18}

\section{Acknowledgments}
The authors wish to thank the Physics Department at the University of Malta for hospitality during the completion of this work.

\end{document}